\documentclass[11pt]{article}

\usepackage[final]{acl}

\usepackage{times}
\usepackage{latexsym}
\usepackage{colortbl}
\usepackage{placeins}

\usepackage[T1]{fontenc}

\usepackage[utf8]{inputenc}

\usepackage{microtype}
\usepackage{textcomp}

\usepackage{inconsolata}

\usepackage{graphicx}
\usepackage{amssymb}
\usepackage{booktabs}
\usepackage{multirow}
\usepackage{tabularx}
\usepackage{etoc}
\usepackage{amsmath}
\usepackage{hyperref}
\usepackage{dblfloatfix}

\newcommand{\eg}{\textit{e.g., }}

\usepackage{xcolor}

\title{Where Does the Sound Go? Tracing Acoustic Information Loss in Audio-Conditioned LLMs}

\author{
  \textbf{Song-ha Jo\textsuperscript{1}\footnotemark[1]},
  \textbf{Sehyun Lee\textsuperscript{2}\footnotemark[1]},
  \textbf{Soyoon Kim\textsuperscript{3}},
  \textbf{Jaesik Choi\textsuperscript{2}},
  \textbf{Sanghyuk Choi\textsuperscript{3}\footnotemark[2]}
\\
\\
  \textsuperscript{1}Seoul National University,
  \textsuperscript{2}KAIST,
  \textsuperscript{3}NAVER Cloud
\\
  \small{
    \texttt{jos02@europa.snu.ac.kr}, \texttt{\{sehyun.lee, jaesik.choi\}@kaist.ac.kr}
  }
\\
  \small{
    \texttt{\{soyoon.kim, sanghyuk.choi\}@navercorp.com}
  }
}

\begin{document}
\maketitle
\renewcommand{\thefootnote}{\fnsymbol{footnote}}
\setcounter{footnote}{1}\footnotetext{\ Equal contribution. This work was done during the residency program at NAVER Cloud.}
\setcounter{footnote}{2}\footnotetext{\ Corresponding author.}
\setcounter{footnote}{0}
\renewcommand{\thefootnote}{\arabic{footnote}}
\addtocontents{toc}{\protect\setcounter{tocdepth}{-3}}
\begin{abstract}
Audio-conditioned language models often underuse acoustic cues such as prosody, emotion, and non-speech sounds, raising the question of whether ASR-supervised frontends discard this information before it reaches the LM. We test whether the frontend is responsible by comparing Whisper-Tiny and Whisper-Small with EnCodec, DAC-VAE, and WavTokenizer in a shared Qwen3.5-4B audio-LM pipeline on ASR, emotion recognition, and sound captioning. Encoder replacement alone does not resolve this underuse: Whisper variants remain strongest overall, including on emotion and environmental sound captioning. To localize the failure, we trace task-relevant information through the encoder, projector, LM layers, and LM head. Linear probes and geometric analyses show that discriminative acoustic structure remains recoverable at the final LM layer, even when MCQA accuracy trails probe accuracy by up to 83 points. Because the answer format and decoding procedure are controlled, this task-dependent gap points to content-specific readout failure rather than generic format bias. LogitLens analyses and a targeted LM head intervention support the conclusion that acoustic underuse is not explained solely by encoder-side information loss and that readout alignment can be a dominant bottleneck.
\end{abstract}
\section{Introduction}

Audio-conditioned language models (audio-LLMs) couple a pretrained audio frontend with a large language model (LM) and have rapidly advanced on speech and audio tasks \cite{QA23,Q2A24,GPT4o24}.
Yet recent benchmarks reveal a persistent limitation: these models often appear to ``transcribe'' rather than ``listen,'' relying on transcript-like content while underusing acoustic cues such as prosody, emotion, speaker style, and non-speech sound attributes \cite{LSTN26}.
This gap between lexical competence and broader acoustic understanding is now well documented, yet its origin within the audio-LLM pipeline is less clear.

\begin{figure}[!t]
    \centering
    \includegraphics[width=\linewidth]{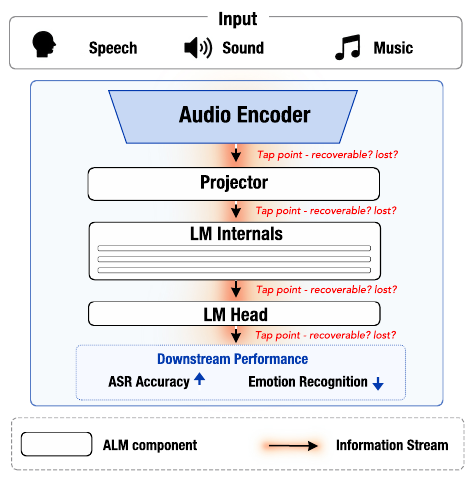}
    \caption{\textbf{Overview of our analysis.} An audio-conditioned LLM (ALM) processes the audio input through an encoder–projector–LM pipeline. To trace where the acoustic information goes, we investigate each stage, after the encoder, projector, LM internals, and LM head, asking whether discriminative acoustic information remains recoverable or has been lost.}
    \label{fig:claim_figure}
\end{figure}

Since many audio-LLMs adopt ASR-supervised encoders, typically Whisper \cite{Wspr23,Q3ASR26}, whose pretraining objective may favor lexical invariance over paralinguistic detail, a natural first suspect is the audio frontend.
Consistent with this intuition, Whisper representations have been observed to underperform self-supervised alternatives such as WavLM on speaker- and emotion-related probing tasks \cite{Enhancing25}, leaving open the possibility that observed acoustic underuse originates before the LM sees the signal.
We therefore treat encoder replacement as a diagnostic counterfactual rather than as an assumed solution.
Reconstruction-based neural audio codecs such as EnCodec, DAC-VAE, and WavTokenizer \cite{ECD22,DAC23,WT25} are trained to reconstruct waveforms, making them a useful test of the encoder-side loss hypothesis: if ASR-supervised frontends are the main bottleneck, representations designed to retain waveform-level detail should at least alleviate acoustic underuse.

We run this test and find that encoder replacement alone is not sufficient in our setting.
We compare Whisper-Tiny and Whisper-Small with three reconstruction-based audio representations from EnCodec, DAC-VAE, and WavTokenizer under a shared Qwen3.5-4B LM \cite{QW35}, each paired with its own projector trained identically, on ASR, emotion recognition, and sound captioning.
Whisper variants \emph{match or outperform} reconstruction-based encoders on emotion recognition and sound captioning, in addition to leading on ASR.
This negative result does not imply that reconstruction-based encoders lack acoustic detail; rather, it shows that the underuse observed downstream cannot be explained simply as acoustic information discarded by ASR-supervised frontends.

If encoder replacement alone does not resolve the gap, where does acoustic information go?
We trace task-relevant information through four stages of the audio-LLM pipeline: the encoder, the projector, the LM's internal layers, and the LM head $W_U$.
Using layer-wise linear probes and complementary geometric analyses with identical targets at each stage, we find that discriminative acoustic structure is not erased by the encoder--projector--LM stack: it remains linearly recoverable at the LM's final hidden states, even when final MCQA predictions are weak.
The probe-to-MCQA gap is highly task-dependent under an identical prompt template, letter set, and decoding procedure, ranging from near zero on CREMA-D to 83 pp on NSynth.
This points to a content-specific readout problem rather than a generic formatting bias.

We test the readout hypothesis causally by fine-tuning only the choice-letter rows of $W_U$ while freezing the encoder, projector, and LM internals.
Across the six analysis tasks, this update touches only $\sim$10 to 26k scalars per task and recovers 30 to 42 pp of MCQA accuracy per encoder on average.
By contrast, adapting the LM internals with LoRA yields gains that are limited and vary with the encoder rather than a broad improvement.
Together, these results show that, in this controlled setting, readout alignment can be the dominant bottleneck even when acoustic evidence survives in the representation.

Taken together, our study makes three main contributions.
First, it tests a natural encoder-side explanation for acoustic underuse under a controlled shared-LM setup, showing that replacing ASR-supervised Whisper with reconstruction-based representations is not sufficient to improve emotion recognition or sound captioning.
Second, it localizes the failure across the encoder--projector--LM pipeline with matched probing and geometric diagnostics, showing that task-relevant acoustic structure remains recoverable up to the final LM hidden states.
Third, it provides readout-level causal evidence by freezing the entire stack and updating only the choice-letter rows of $W_U$, and uses LogitLens and alignment analyses to characterize how otherwise recoverable acoustic evidence fails to surface as the correct verbalized prediction.
These results reframe acoustic underuse, at least in our controlled setting, as a failure that can arise at readout rather than from encoder-side acoustic loss alone, and identify LM-head alignment as a useful diagnostic lens for understanding such failures.

\begin{table}[!t]
\centering
\small
\setlength{\tabcolsep}{4pt}
\resizebox{\columnwidth}{!}{%
\begin{tabular}{@{}lllrr@{}}
\toprule
Encoder & Arch. & Obj. & Tokens/s & Params \\
\midrule
EnCodec       & Conv      & Recon.       & $75$ & $7.43$M  \\
DACVAE        & Conv+VAE  & Recon. + WM  & $25$ & $27.55$M \\
Whisper-Small & Tr        & ASR          & $50$ & $88.15$M \\
Whisper-Tiny  & Tr        & ASR          & $50$ & $8.21$M  \\
WavTokenizer  & Conv      & Recon.       & $40$ & $8.30$M  \\
\bottomrule
\end{tabular}%
}
\caption{Audio encoder candidates. Arch.: Conv (SEANet-style stack), Tr (Transformer), VAE (variational bottleneck). Obj.: Recon. (reconstruction-based, \eg neural codec), Recon.\,+\,WM (with acoustic watermarking), ASR (supervised speech recognition). Params count the active forward path. Per-encoder sampling rate, hop, latent dimensionality, pretrained URIs, and training-time settings are listed in Table \ref{tab:training-config}.}
\label{tab:encoders}
\end{table}

\section{Related Work}

\paragraph{Audio frontends in audio-LLMs.}
Audio-conditioned language models typically connect an audio frontend to a text-trained language model through a learned adapter or projector.
Many recent systems use transcription-oriented frontends, with Whisper \cite{Wspr23} widely adopted in Qwen-Audio \cite{QA23}, Qwen2-Audio \cite{Q2A24}, and Audio Flamingo 3 \cite{AudioFlamingo3_25}.
These encoders are highly effective for lexical tasks, but their ASR-oriented objectives may encourage invariance to speaker, prosodic, and environmental variation.
Consistent with this concern, Whisper representations have been reported to underperform self-supervised alternatives such as WavLM on speaker- and emotion-related probing tasks \cite{Enhancing25}.
This makes the audio frontend a plausible source of acoustic underuse, but does not establish whether replacing it is sufficient.

Several systems therefore complement or depart from transcription-oriented frontends.
SALMONN \cite{SALMONN24} combines Whisper with BEATs \cite{BEATS23} in a dual-encoder architecture, while codec-based models such as Moshi \cite{MOSHI24} show that reconstruction-oriented audio representations can support spoken language modeling.
Neural audio codecs and tokenizers such as EnCodec \cite{ECD22}, DAC-VAE \cite{DACVAE24}, and WavTokenizer \cite{WT25} are trained to reconstruct audio and therefore provide a useful contrast to ASR-supervised representations when testing whether acoustic underuse originates at the frontend.
Our work compares these frontend families under a shared LM setup, not to assume that reconstruction-based encoders should dominate, but to test whether frontend replacement explains the acoustic-underuse behavior observed in audio-LLMs.

\paragraph{Acoustic underuse and representation analysis.}
Recent work has shown that strong audio-LLM performance does not necessarily imply reliance on acoustic evidence.
Most directly, \citet{LSTN26} introduce LISTEN, a benchmark that separates lexical and acoustic emotion cues, and show that contemporary audio-LLMs often follow transcript-like content even when acoustic cues indicate a different emotion.
This behavioral finding motivates our analysis, but it does not localize where the relevant information is lost or suppressed: acoustic cues may be weakened by the frontend, discarded by the projector, transformed inside the LM, or present in hidden states but not expressed by the final LM head.

Analytical tools such as linear probing \cite{Layerwise21,Probing23} and representation similarity analyses like centered kernel alignment \cite{CKA19} have been used to inspect speech and audio representations.
However, these studies typically focus on a single encoder and stop at the encoder output.
Recent efforts have begun to examine encoder--LM interfaces, including audio representation transfer analyses \cite{PAL25} and unified audio-encoder benchmarks for ALMs \cite{IS26Chal}.
Our work extends this diagnostic view across the full encoder--projector--LM pipeline, using identical probing targets at each stage to distinguish representation-level loss from later readout failure.

\paragraph{Readout bias in verbalized classification.}
Language-model predictions are mediated by the LM head and can be sensitive to label words, answer formats, and multiple-choice options.
In text classification, this issue is commonly studied through verbalizers and prompt-based classification, where different surface forms for the same class can substantially change model predictions \cite{SchickSchutze21PET,LiuEtAl23PromptSurvey}.
Similar concerns arise in multiple-choice evaluation, where option letters and formatting can affect predictions independently of the underlying evidence \cite{ZhengEtAl24MCQBias}.
For audio-LLMs, this creates a possible gap between information that is recoverable from hidden states and information that is expressed as final answer tokens.
Our work connects this readout problem to acoustic underuse by showing that task-relevant acoustic structure remains recoverable in the final LM layer, while the LM head fails to map that structure to the correct choice-letter tokens.

\section{Experiment Design}

\label{sec:methodology}

\subsection{Architecture}
\label{sec:method:arch}

We adopt a three-stage encoder-projector-LM architecture \cite{FLM25,SLAM24}, where a frozen audio encoder maps a waveform to a continuous latent sequence, a trainable projector aligns this latent with the LM embedding space, and a pretrained LM autoregressively generates a text response.

\paragraph{Audio encoders.}
Most ALMs default to an ASR-based encoder, typically Whisper, yet evaluations like LISTEN suggest acoustic detail is underutilized downstream.
To test whether this acoustic information bottleneck originates in the encoder, we treat the audio encoder as a swappable component and study five pretrained candidates spanning both encoder families (reconstruction-based and ASR-based), summarized in Table \ref{tab:encoders}.
Three are reconstruction-based encoders, namely \textbf{EnCodec} \cite{ECD22}, \textbf{DAC-VAE} \cite{DAC23,AS24}, and \textbf{WavTokenizer} \cite{WT25}, while the remaining two are ASR-based encoders, \textbf{Whisper-Small} and \textbf{Whisper-Tiny} \cite{Wspr23}.
For the reconstruction-based encoders, we extract only the feed-forward encode path, omitting any decoder, residual vector quantizer, watermarker, or output projection.
For the ASR-based encoders, we use the encoder stack only, discarding the text decoder.
Self-supervised encoders fall outside this transcription-versus-reconstruction contrast and are therefore not part of this pool.
In all cases, the encoder is frozen throughout overall training.

\paragraph{Trainable components.}
We first train only the projector, a four-layer causal Transformer decoder, and freeze everything else.
To test whether the LM degrades the acoustic representation passed from the projector, we then fine-tune the LM (Qwen3.5-4B \cite{QW35}) through low-rank adapters \cite{Hu22} while continuing to update the projector.
We use Qwen3.5-4B throughout the main text, and Appendix~\ref{app:second_lm} repeats the analysis on a different LM, Ministral3-3B (Instruct).
Training uses a multi-task mixture of English ASR, environmental-sound captioning, and emotion recognition.
The projector architecture, LM configuration, training-data mixture, and optimization settings are all detailed in Appendix~\ref{app:impl}.

\subsection{Analysis Settings and Methods}
\label{sec:method:analysis}

\paragraph{Evaluation benchmark.}\label{sec:method:eval}
We evaluate on a multi-task suite spanning ASR, emotion, and sound captioning (Table~\ref{tab:encoder_comparison}).
ASR uses the LibriSpeech \cite{librispeech} and GigaSpeech \cite{gigaspeech} with WER and CER after Whisper-style English text normalization  \cite{Wspr23}.
Emotion uses MELD \cite{meld} and a leakage-free IEMOCAP \cite{iemocap} Session 5 split with accuracy and macro-F1.
Sound captioning covers FSD50K \cite{fsd50k}, AudioSet \cite{audioset}, AudioCaps \cite{audiocaps}, and Clotho \cite{clotho} with BLEU-1, BLEU-4, and CIDEr \cite{cider}.
Decoding is greedy throughout (\texttt{num\_beams}$=1$, \texttt{do\_sample}$=$False), with no external language model, and all encoder variants share identical evaluation code, tokenization, and normalization.

\section{Analysis}

\subsection{Acoustic Underuse Persists across Encoders}
ALMs underuse acoustic cues despite using competent audio encoders \cite{LSTN26}.
This phenomenon raises the question of what part of the pipeline is responsible, and a first hypothesis blames the encoder training objective.
Although ASR supervision may suppress non-lexical structure even when the encoder is otherwise capable, the dominant choice of ASR-supervised Whisper has not been systematically validated.
To test this hypothesis, we compared five encoders, spanning ASR-based and reconstruction-based training, with end-task performance.

The cross-benchmark comparison in Table \ref{tab:results-summary} shows clear differences, confirming that encoder choice strongly affects downstream behavior.
Whisper-Small leads on every task family on average, followed by Whisper-Tiny, and this lead extends beyond ASR to emotion and sound captioning.
Even the best-performing encoder, however, leaves substantial headroom on emotion and captioning, so the acoustic underuse persists regardless of encoder choice.
Task-level differences across encoders therefore reflect a relative ranking rather than a localization of where the failure originates inside the pipeline.
Because the encoders differ in training data and architecture, we treat the ranking as descriptive and establish the localization within each encoder independently.

To begin localizing the bottleneck, we compare encoder-side linear probes against final LM-side prediction accuracy (MCQA format) on the same tasks (Table \ref{tab:audio_benchmark_summary}).
Across domains, probes often outperform MCQA substantially, indicating that task-relevant information is recoverable from encoder latents even when the LM's prediction is weak.
The bottleneck therefore lies after the encoder, motivating the stage-wise analyses below.

\begin{table}[t]
\centering
\small
\setlength{\tabcolsep}{4pt}
\resizebox{\columnwidth}{!}{%
\begin{tabular}{lrrr}
\toprule
\textbf{Encoder}       & \textbf{ASR}     & \textbf{Emotion}        & \textbf{Captioning} \\
                       & WER $\downarrow$ & macro-F1 $\uparrow$     & CIDEr $\uparrow$ \\
\midrule
Whisper-Small & $\mathbf{7.05}$      & $\mathbf{0.462}$            & $\mathbf{0.847}$ \\
Whisper-Tiny  & $\underline{8.84}$   & $\underline{0.444}$         & $\underline{0.685}$ \\
EnCodec       & $20.80$              & $0.393$                     & $0.259$ \\
WavTokenizer  & $33.61$              & $0.356$                     & $0.325$ \\
DAC-VAE       & $20.46$              & $0.400$                     & $0.235$ \\
\bottomrule
\end{tabular}%
}
\caption{Per-encoder summary across task families, averaged within each family. ASR WER averages LibriSpeech test-clean, test-other, and GigaSpeech test. Emotion macro-F1 averages MELD and IEMOCAP-S5. Captioning CIDEr averages FSD50K, Clotho, AudioCaps, and AudioSet. Per-benchmark detail is in Table \ref{tab:encoder_comparison}.}
\label{tab:results-summary}
\end{table}

\begin{table*}[t]
\centering
\small
\setlength{\tabcolsep}{5pt}
\begin{tabular}{l ccccc}
\toprule
\textbf{Task} & \textbf{Whisper-Tiny} & \textbf{Whisper-Small} & \textbf{Encodec} & \textbf{WavTokenizer} & \textbf{DAC-VAE} \\
\midrule
CREMA-D       & 57/31 ($-25$) & 59/28 ($-31$) & 30/30 ($+0$)  & 34/22 ($-12$) & 29/22 ($-7$)  \\
IEMOCAP       & 65/53 ($-12$) & 69/58 ($-11$) & 56/52 ($-5$)  & 52/44 ($-8$)  & 52/45 ($-6$)  \\
SpeechCmds    & 97/83 ($-14$) & 98/66 ($-31$) & 49/41 ($-7$)  & 25/47 ($+22$) & 35/44 ($+9$)  \\
AudioMNIST    & 100/88 ($-12$) & 100/53 ($-47$) & 85/17 ($-68$) & 64/14 ($-50$) & 77/10 ($-67$) \\
VocalSound    & 83/41 ($-42$) & 88/52 ($-36$) & 55/27 ($-29$) & 35/29 ($-6$)  & 44/26 ($-18$) \\
ESC-10        & 93/37 ($-56$) & 95/50 ($-45$) & 76/44 ($-33$) & 62/20 ($-42$) & 66/20 ($-47$) \\
ESC-50 (5cat) & 79/33 ($-46$) & 89/43 ($-47$) & 53/38 ($-15$) & 41/23 ($-18$) & 45/20 ($-26$) \\
UrbanSound8K  & 86/26 ($-60$) & 90/31 ($-59$) & 70/20 ($-50$) & 34/15 ($-19$) & 52/9 ($-43$)  \\
TAU           & 58/8 ($-50$)  & 59/10 ($-49$) & 59/11 ($-48$) & 41/12 ($-30$) & 55/10 ($-44$) \\
GTZAN         & 82/9 ($-73$)  & 84/15 ($-69$) & 66/11 ($-55$) & 47/10 ($-37$) & 58/9 ($-49$)  \\
NSynth        & 90/8 ($-83$)  & 97/14 ($-83$) & 61/17 ($-44$) & 38/10 ($-27$) & 56/7 ($-49$)  \\
IRMAS         & 54/11 ($-43$) & 66/12 ($-53$) & 44/10 ($-34$) & 32/11 ($-21$) & 40/12 ($-28$) \\
MedleySolos   & 92/24 ($-67$) & 96/26 ($-71$) & 89/18 ($-71$) & 63/18 ($-44$) & 80/17 ($-63$) \\
FSD50K\textsuperscript{$\dagger$}        & 82/47 ($-35$) & 82/57 ($-26$) & 72/44 ($-27$) & 63/38 ($-26$) & 68/32 ($-36$) \\
AudioSet\textsuperscript{$\dagger$}      & 75/47 ($-28$) & 77/56 ($-21$) & 71/42 ($-29$) & 63/30 ($-33$) & 63/29 ($-33$) \\
\midrule
\textit{Macro avg.}   & 79.5/36.4 ($-43.1$) & 83.3/38.1 ($-45.2$) & 62.4/28.1 ($-34.3$) & 46.3/22.9 ($-23.4$) & 54.7/20.8 ($-33.9$) \\
\bottomrule
\end{tabular}
\caption{Probe / MCQA accuracy (\%) per task across encoders, with the probe$\to$MCQA transfer gap in parentheses. Tasks span speech emotion (CREMA-D, IEMOCAP), speech content (SpeechCmds, AudioMNIST), vocal sounds (VocalSound), environmental sound (ESC-10, ESC-50 (5cat), UrbanSound8K, TAU, FSD50K, AudioSet), and music (GTZAN, NSynth, IRMAS, MedleySolos); per-task class count, balance, and chance level are listed in Table~\ref{tab:audio_benchmark_meta}. \textsuperscript{$\dagger$}FSD50K (200 cls, $n{=}10{,}231$) and AudioSet (632 cls, $n{=}3{,}429$ aligned) use 4-way MCQA with random distractors (chance $25\%$), since per-task $K$-way is impractical at that class count.}
\label{tab:audio_benchmark_summary}
\end{table*}

\subsection{Acoustic Information Survives Encoder, Projector, and LM Internals}
If encoder choice alone does not explain the downstream acoustic gap, the failure must originate somewhere later in the pipeline.
A natural starting point is that information is degraded at the encoder--LM interface, when the projector compresses encoder latents into the LM's text-trained representation space.
We test this hypothesis by tracking the recoverability of acoustic structure at every stage of the pipeline -- encoder, projector, and LM -- using three diagnostics with shared targets: (1) linear-probe accuracy for predictive recoverability, (2) a within-/across-class distance ratio for geometric preservation, and (3) LM-side adaptability experiments.

\begin{figure*}[t]
  \centering
  \includegraphics[width=\textwidth]{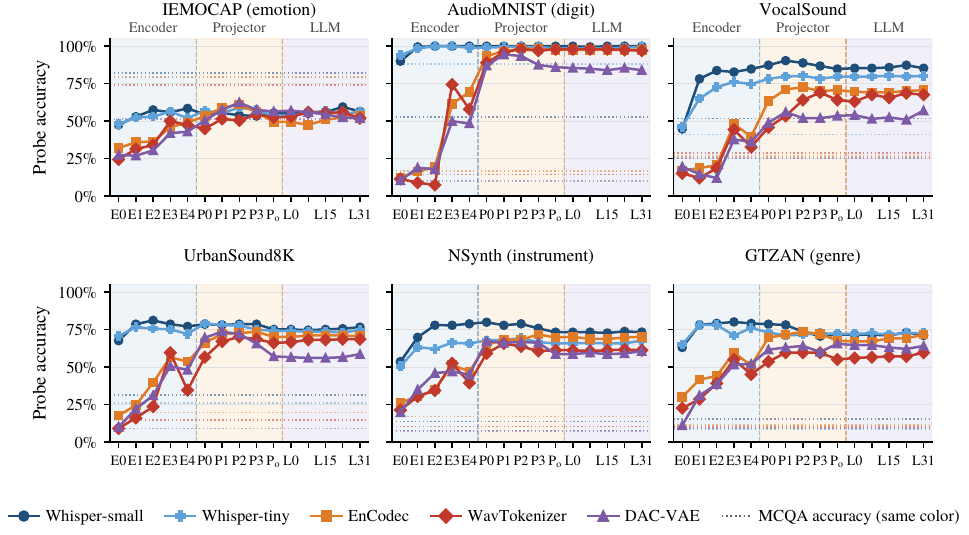}
  \caption{Layer-wise probe accuracy across encoder ($E0$--$E4$), projector ($P0$--$P3$, $P_o$), and LM ($L0$, $L8$, $L15$, $L23$, $L31$) stages. Dashed horizontal lines indicate final LM MCQA accuracy for the same encoder color.}
  \label{fig:layer_probe_grid}
\end{figure*}

\paragraph{(1) Layer-wise linear probing.}
As shown in Figure~\ref{fig:layer_probe_grid}, linear-probe accuracy is measured at taps spanning the pipeline: five encoder taps ($E0$--$E4$), four projector taps plus the projector output ($P0$--$P3$, $P_o$), and five LM taps ($L0$, $L8$, $L15$, $L23$, $L31$ of the 32-layer LM), with an identical probe fit at every tap.
The encoder-to-projector transition consistently increases probe recoverability on non-ASR acoustic tasks rather than degrading it, and this gain is largely retained through LM depth instead of collapsing at later layers.
Probe trajectories therefore remain substantially above the corresponding final-LM MCQA accuracy lines across datasets, indicating that discriminative acoustic structure survives well past the interface boundary.
The probe is fitted separately from the model, however, so this establishes only that the information is present in the hidden states, not that the model routes it to the correct answer.

\begin{figure}[t]
  \centering
  \includegraphics[width=\columnwidth]{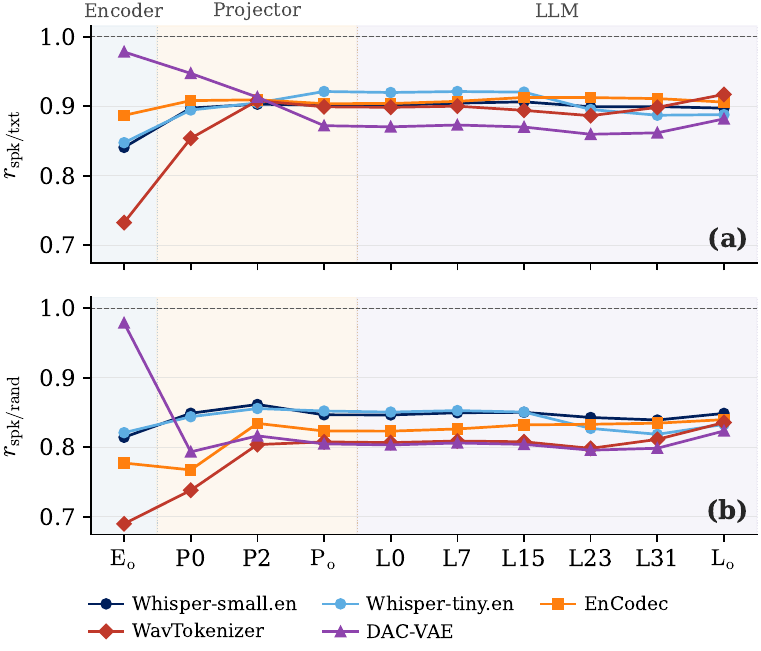}
  \caption{Per-stage distance ratios across the five encoders on IEMOCAP Session~5. \textbf{(a)} Across-/within-speaker distance ratio (same transcript, different speaker $/$ same speaker, different transcript). \textbf{(b)} Across-speaker $/$ random-pair distance ratio (same transcript, different speaker $/$ different speaker, different transcript).}
  \label{fig:distance_ratio}
\end{figure}

\paragraph{(2) Within-/across-speaker geometry.}
To verify that this survival reflects the preserved \emph{discriminative} structure rather than merely co-located features, Figure~\ref{fig:distance_ratio} reports two complementary distance ratios across encoder, projector, and LM stages on IEMOCAP Session~5, where speaker and transcript labels are controlled.
For an utterance pair $(i, j)$ with stage representations $h_i, h_j$, speaker labels $s_i, s_j$, and transcripts $t_i, t_j$, the mean cosine distance over a pair set $P$ is
\[
\overline{d}(P) = \frac{1}{|P|} \sum_{(i,j) \in P} 1 - \frac{h_i^{\top} h_j}{\lVert h_i \rVert\, \lVert h_j \rVert},
\]
with the three pair sets defined as $P_{\text{spk}} = \{(i,j) : t_i = t_j,\, s_i \neq s_j\}$, $P_{\text{txt}} = \{(i,j) : s_i = s_j,\, t_i \neq t_j\}$, and $P_{\text{rand}} = \{(i,j) : s_i \neq s_j,\, t_i \neq t_j\}$.
Figure~\ref{fig:distance_ratio}(a) measures speaker variation relative to transcript variation,
\[
r_{\text{spk/txt}} = \frac{\overline{d}(P_{\text{spk}})}{\overline{d}(P_{\text{txt}})},
\]
and Figure~\ref{fig:distance_ratio}(b) measures it relative to the random-pair baseline,
\[
r_{\text{spk/rand}} = \frac{\overline{d}(P_{\text{spk}})}{\overline{d}(P_{\text{rand}})}.
\]

In both metrics, higher values indicate stronger preservation of speaker-specific geometry relative to the comparison baseline.
A value of $r_{\text{spk/txt}} = 1$ marks the balance point where speaker variation produces the same average distance as transcript variation, so the representation places its acoustic and semantic axes on equal footing.
Values below $1$ indicate transcript-dominated geometry (the semantic axis stronger than the acoustic one), and values above $1$ indicate the reverse.
A value of $r_{\text{spk/rand}} = 1$ marks the point where pure speaker variation accounts for the full random-pair distance, so the representation treats two utterances of the same transcript by different speakers as no closer than a fully random pair.
Values below $1$ mean that controlling the transcript reduces the distance, so transcript variation contributes additional separation beyond speaker variation alone.

\begin{table*}[t]
\centering
\small
\setlength{\tabcolsep}{4pt}
\begin{tabular}{l rrr rrr rrr}
\toprule
 & \multicolumn{3}{c}{\textbf{ASR} (WER\,$\downarrow$)}
 & \multicolumn{3}{c}{\textbf{Emotion} (macro-F1\,$\uparrow$)}
 & \multicolumn{3}{c}{\textbf{Captioning} (CIDEr\,$\uparrow$)} \\
\cmidrule(lr){2-4}\cmidrule(lr){5-7}\cmidrule(lr){8-10}
\textbf{Encoder} & align & FT & $\Delta$ & align & FT & $\Delta$ & align & FT & $\Delta$ \\
\midrule
Whisper-Tiny & $8.84$  & $6.06$  & $-2.78$         & $0.444$ & $0.431$ & $-0.013$         & $0.685$ & $0.476$ & $-0.209$ \\
EnCodec      & $20.80$ & $10.65$ & $\mathbf{-10.15}$ & $0.393$ & $0.402$ & $+0.009$         & $0.259$ & $0.408$ & $\mathbf{+0.149}$ \\
DAC-VAE      & $20.46$ & $16.60$ & $-3.86$         & $0.400$ & $0.382$ & $-0.018$         & $0.235$ & $0.344$ & $+0.109$ \\
\bottomrule
\end{tabular}
\caption{Family-level effect of LoRA fine-tuning (FT) over align, averaged within each task family using the same metric pool as Table \ref{tab:results-summary}. $\Delta$ is FT $-$ align; for ASR, negative $\Delta$ is improvement. Per-task numbers and the best-by-rank-sum FT checkpoint per encoder are in Appendix Table \ref{tab:lora_per_task}.}
\label{tab:lora-summary}
\end{table*}

All observed values lie below $1$, placing every encoder in the transcript-dominated regime, yet they stay well above the collapse range, so speaker identity remains preserved across the pipeline.
Within this regime, reconstruction-based DAC-VAE starts highest at $r_{\text{spk/txt}} = 0.978$ at the encoder output and gradually compresses through the projector and LM to $0.882$ at $\mathrm{lm.out}$.
ASR-based Whisper encoders start a little lower at the encoder output ($0.841$ for Whisper-Small, $0.848$ for Whisper-Tiny), but still meaningful.
These values rise in the projector, and settle near $0.89$ through the LM ($0.897$ and $0.888$ at $\mathrm{lm.out}$, respectively), showing that even an ASR loss leaves enough speaker structure at the encoder output for the projector and LM to preserve and propagate through depth.

\paragraph{(3) Encoder-wise LoRA fine-tuning effect and alignment potential.}
A complementary check is whether targeted LM modification can close the audio-task gap.
We fine-tune the LM via LoRA (rank 32, $\alpha{=}64$ on the seven linear projections per Transformer block, attention $q/k/v/o$ and MLP $gate/up/down$) on top of the frozen encoder and a co-fine-tuned projector.
To isolate the LoRA contribution, we zero only the LoRA $B$ matrices while keeping the projector intact and re-evaluate the nine downstream tasks.

Across encoders, the LoRA effect varies in both magnitude and sign (Table \ref{tab:lora-summary}), in line with each encoder's alignment potential rather than any single LM-internal circuit.
By alignment potential we mean how much further a representation could be aligned to the LM, a property of the representation itself rather than a quantity we read off directly.
The align-stage score is only a rough proxy for it.
The encoder with the strongest align baseline (Whisper-Tiny) shows only modest ASR gains, near-flat emotion, and a captioning regression, while the encoders with weak align baselines (EnCodec, DAC-VAE) recover substantial captioning headroom and, for EnCodec, substantially reduce ASR WER.
Per-encoder details are in Appendix \ref{app:lora_details}.

\begin{figure}[t]
  \centering
  \includegraphics[width=\columnwidth]{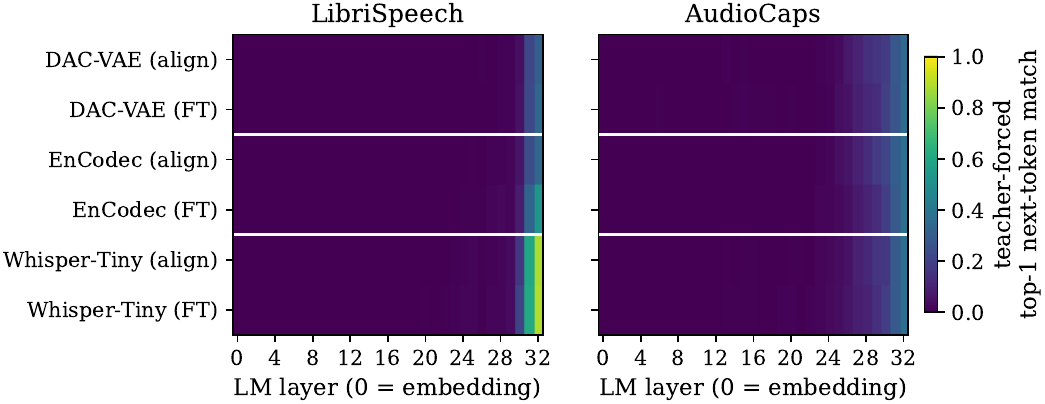}
  \caption{Layer-wise LogitLens heatmap of teacher-forced top-1 next-token match across the 32 LM blocks plus the input embedding ($x = 0$), for three encoders (Whisper-Tiny / EnCodec / DAC-VAE) in two conditions (align, fine-tuning).}
  \label{fig:logit_lens}
\end{figure}

The LogitLens view of LM-internal decoding is consistent with the LoRA results (Figure \ref{fig:logit_lens}).
On LibriSpeech, Whisper-Tiny's final-layer top-1 accuracy moves only from $0.873$ align to $0.882$ after fine-tuning, whereas EnCodec moves from $0.331$ to $0.520$, implying substantial alignment potential.
The same pattern holds for AudioCaps, where the flatness across align $\to$ fine-tuning reflects the LM-internal trajectory not gaining additional captioning structure from LoRA.
That the LibriSpeech shift concentrates at the last few LM layers and that neither task shows broad LM-internal change suggests that the gap concentrates in the readout-level alignment rather than in LM internal capacity.

All three probes converge on the same verdict.
$r$ stays below the equality line yet is far from collapse, probe accuracy remains substantially above final-LM MCQA accuracy at every depth, and LoRA fine-tuning leaves LM internal capacity untouched, concentrating its effect at the readout-level alignment.
The encoder--projector--LM stack therefore does not erase acoustic information, shifting the failure hypothesis from \emph{complete information loss} toward \emph{under-utilization at the LM prediction stage} and motivating the readout-level analyses below.

\subsection{The LM Readout, Not Its Representations, Drives MCQA Failure}
\label{sec:readout}
If acoustic information survives the encoder--projector--LM stack, the remaining failure must occur when the LM converts hidden states into a discrete answer letter through the output projection $W_U$.
A natural first hypothesis is that the LM still uses this surviving evidence at prediction time, so audio-conditioned MCQA accuracy should rise above a text-only baseline by an amount that tracks each encoder's audio-task strength.

\begin{figure}[t]
  \centering
  \includegraphics[width=\columnwidth]{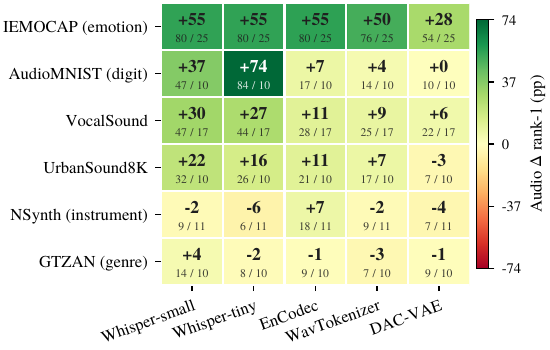}
  \caption{Rank-delta heatmap of the audio effect: the increase in rank-1 letter accuracy (\%) when audio is present versus a text-only baseline. Rows are six tasks; columns are five encoders.}
  \label{fig:rank_delta_heatmap}
\end{figure}

\paragraph{(1) Audio effect.}
The \emph{audio effect} tests this hypothesis: for each sample, a teacher-forced MCQA prompt of the form ``\texttt{\{question\}\textbackslash nChoices: A) $c_1$ \ldots K) $c_K$\textbackslash nAnswer with the letter.}'' yields logits at the final prompt position, restricted to the candidate-letter tokens $\{A,\dots,K\}$; \emph{rank-1 letter accuracy} is the fraction of samples where the gold letter has the highest restricted logit, and the audio effect is the increase in this rank-1 accuracy between the audio-on condition and a text-only baseline that inserts no audio placeholder tokens at all, leaving the rest of the prompt text identical (200-sample subset per task across six tasks, identical prompts across the five encoder families).
As shown in Figure~\ref{fig:rank_delta_heatmap}, a sharp split emerges: speech tasks (emotion, digits, vocal sounds) gain $27$ to $74$\,pp, while music tasks (NSynth, GTZAN) show near-zero or negative gains; however, this split is consistent across encoders, arguing against a purely encoder-side attribution and pointing to the LM's answer mapping as the failure site.

\paragraph{(2) $W_U$ letter-row surgical recovery.}
This answer-mapping failure points to the LM's output projection $W_U$, the layer that converts final hidden states into candidate-letter logits.
To verify this causally, every weight is frozen except the $n_c$ rows of $W_U$ corresponding to the choice letters, and those rows are fine-tuned with cross-entropy on saved $L31$ assistant-newline hidden states, the last position before the answer token (80/20 stratified split, full-batch Adam, lr $10^{-3}$, 200 steps; held-out test accuracy reported).
Unlike a linear probe, this update stays inside the model's own output path.
The remaining vocabulary rows, the LM-head geometry outside the choice letters, and the decoding interface are all left as they are, so the recovered accuracy is produced by the same projection that generates the model's own predictions.
This update of only $n_c \times 2560$ scalars per task ($\approx 10$--$26$k, depending on the class count) recovers $30$ to $42$\,pp of MCQA accuracy on average across the five encoders and six tasks (Figure~\ref{fig:causal_verify}; per-task breakdown in Appendix~\ref{app:wu_finetune_grid}).
To characterize not only \emph{that} $W_U$ fails but also \emph{how} it fails, we report LogitLens decoding in Appendix~\ref{app:logit_lens_grid}.

\begin{figure}[t]
  \centering
  \includegraphics[width=0.85\columnwidth]{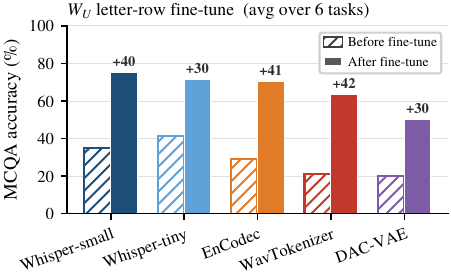}
  \caption{$W_U$ letter-row surgical recovery, averaged over the six analysis tasks. Bars compare MCQA accuracy before and after fine-tuning only the $n_c$ rows of $W_U$ indexing the choice letters ($n_c \times 2560 \approx 10$--$26$k parameters per task), across five encoders; the $+\Delta$ label above each After bar is the per-encoder absolute gain (mean across tasks).}
  \label{fig:causal_verify}
\end{figure}

MCQA failures are therefore driven by readout-level mismatch rather than absence of acoustic evidence.
The LM contains recoverable acoustic structure at $L31$ (probes are accurate), $W_U$ fails to route it toward the correct letter, and a tiny update to its letter rows recovers most of the MCQA gap.

\section{Conclusion}
We asked whether audio-LLMs' underuse of paralinguistic and non-speech information arises from their ASR-supervised encoders, which are often suspected of prioritizing transcription over acoustic detail, or from later stages of the encoder, projector, and LM pipeline. Replacing Whisper with reconstruction-based representations did not improve emotion recognition or sound captioning, showing that a practical frontend swap is not a sufficient remedy. Instead, discriminative acoustic structure remains linearly recoverable at the LM's final hidden states, well above multiple-choice question answering accuracy. The failure lies at the readout: fine-tuning a small subset of $W_U$ rows recovers accuracy across all five encoders, while LoRA adaptation of the LM internals yields gains that are limited and vary with the encoder.

More broadly, our analysis shows the value of localizing where acoustic information is lost rather than assuming the encoder is at fault. Tracing information stage by stage separates what the model fails to encode from what it fails to use, revealing that the evidence is already present and that the limiting factor lies downstream. These findings carry a clear implication for audio-LLM design: effective design follows from diagnosis, by locating the stage that bottlenecks acoustic understanding and targeting it directly.

\section*{Limitations}

While our analysis localizes the dominant failure mode of audio-conditioned MCQA, several dimensions warrant further investigation. Our experiments use two base LMs (Qwen3.5-4B and Ministral3-3B), both small and instruction-tuned, and whether the readout bottleneck persists at larger scales or under different instruction-tuning regimes remains open. Our stage-wise readout analysis covers only MCQA-format classification and our speaker-geometry analysis only IEMOCAP Session 5, though extending the same localization framework to generative tasks such as ASR and free-form sound captioning would give a fuller picture. Our use of off-the-shelf encoders also makes the cross-encoder ranking an end-to-end comparison rather than a controlled attribution, a gap that future work should close by isolating capacity and pretraining differences and adding self-supervised encoders (HuBERT, WavLM, wav2vec~2.0) and semantic-aligned hybrid codecs (Mimi, SpeechTokenizer).

Methodologically, probing establishes that acoustic information is \emph{recoverable} but not that the model \emph{uses} it, so we treat probe accuracy as an upper bound and rely on the causal $W_U$ intervention to bridge the two, while the letter-row update is a post-hoc diagnostic rather than a deployment recipe, and turning it into a training-time method, for example answer-format-aware alignment or label-direction regularizers, is a natural next step. We expect the letter bias we isolate to be a measurable special case of a vocabulary-level readout bias, since similar failures to express available perceptual evidence have been reported in other modalities \cite{Hicke25,Li26,Lee26}, and testing that in open-ended generation is the most promising next step.

\bibliography{custom}
\clearpage

\FloatBarrier
\appendix
\addtocontents{toc}{\protect\setcounter{tocdepth}{2}}
\setcounter{tocdepth}{2}
\section*{Appendix}
\noindent\rule{\columnwidth}{0.4pt}\vspace{-2pt}
\renewcommand{\contentsname}{}
\etocsettocstyle{}{}
\etocsetnexttocdepth{subsection}
\tableofcontents
\noindent\rule{\columnwidth}{0.4pt}\vspace{4pt}

\section{Implementation Details}
\label{app:impl}

\subsection{Model}
\label{app:model}

\paragraph{Pretrained checkpoints.}
\label{app:checkpoints}
Table \ref{tab:training-config} lists the Hugging Face repository identifiers of the pretrained audio encoders used in this work, together with their per-encoder specifications and training-time settings.

\begin{table*}[t]
\centering
\small
\setlength{\tabcolsep}{4pt}
\resizebox{\textwidth}{!}{%
\begin{tabular}{llrrrrrr}
\toprule
Encoder & HF repository & SR (kHz) & Hop & Dim. & cutoff (tok) & Global batch & Rows / phase \\
\midrule
Whisper-Small  & \texttt{openai/whisper-small.en}                  & $16$ & $320$  & $768$ & $3584$ & $32$ & ${\sim}24.4$M \\
Whisper-Tiny   & \texttt{openai/whisper-tiny.en}                   & $16$ & $320$  & $384$ & $3584$ & $32$ & ${\sim}24.4$M \\
EnCodec        & \texttt{facebook/encodec\_24khz}                  & $24$ & $320$  & $128$ & $4096$ & $32$ & ${\sim}27.9$M \\
WavTokenizer   & \texttt{novateur/WavTokenizer-large-unify-40token} & $24$ & $600$  & $512$ & $3584$ & $32$ & ${\sim}24.4$M \\
DAC-VAE        & \texttt{facebook/dacvae-watermarked}              & $48$ & $1920$ & $128$ & $3584$ & $16$ & ${\sim}12.2$M \\
\bottomrule
\end{tabular}%
}
\caption{Per-encoder spec, pretrained checkpoint, and training-time settings. Dim. is the encoder latent dimensionality (projector input). Global batch $=$ $8$ (GPUs) $\times$ per-device batch $\times$ gradient accumulation. Rows / phase is computed from $100{,}000$ steps, global batch, cutoff length, and the row-size weighted mean of $\approx 470$ tokens per training row. EnCodec uses a larger cutoff ($4096$ tokens) because its $75$ tokens/s rate fills $2{,}250$ audio frames for a $30$-second utterance and exceeds $3584$ once the ChatML wrapper is added; DAC-VAE uses a halved global batch ($16$) because its $48$\,kHz input carries roughly $3\times$ as many raw audio samples per utterance as the $16$/$24$\,kHz encoders, forcing a smaller per-step batch under fixed GPU memory.}
\label{tab:training-config}
\end{table*}

\paragraph{Projector.}
\label{app:projector}
The projector has hidden size $512$, $8$ attention heads, and a feed-forward dimension of $2048$.
Inside the projector, an input linear projection $\mathbb{R}^{d_{\text{enc}}} \!\to\! \mathbb{R}^{512}$ feeds the four decoder layers, whose output is mapped to $\mathbb{R}^{2560}$ by an output linear projection, where $d_{\text{enc}}$ is the encoder's latent dimensionality listed in Table \ref{tab:training-config}.
Each decoder layer uses pre-norm RMSNorm \cite{ZL20}, SwiGLU \cite{Sh20}, rotary position encoding \cite{SP23}, FlashAttention-2 causal attention \cite{Dao23}, and bias-free linear projections throughout.
The four decoder layers, the final RMSNorm, and the output projection are shared across all encoders and contribute $\sim\!18$M parameters.
Only the $d_{\text{enc}} \!\times\! 512$ input projection varies with the choice of encoder, leaving the total projector size at roughly $18$M parameters across all configurations (about $0.45\%$ of the LLM).

\paragraph{Language model.}
\label{app:lm}
We obtain our LM from Qwen3.5-4B \cite{QW35} solely by extending the tokenizer with reserved audio control tokens that demarcate where projector outputs are spliced into the input.
The transformer architecture and pretrained weights are inherited unchanged, so audio conditioning is injected at the input-embedding level rather than through cross-attention or in-network adapters.

\subsection{Training}
\label{app:training}

\paragraph{Per-modality corpora.}
\label{app:data}
The ASR portion combines GigaSpeech XL (50\% subsample), LibriTTS-R \cite{Ko23}, the English subset of Multilingual LibriSpeech (MLS) \cite{Pr20}, and the English portion of VoxPopuli \cite{Wa21}.
The environmental-sound portion aggregates the LAION-Audio-630k collections (Freesound, Epidemic, BBC, Audiostock), AudioCaps, FSD50K, the AudioSet balanced split, Clotho, and MACS.\footnote{Because these collections are independently curated from overlapping public sources, we cannot fully rule out intra-LAION overlap (\eg between Freesound, BBC, Epidemic, and Audiostock) or inter-corpus overlap among the other sound-captioning datasets. The LAION-Audio-630k authors documented concrete cross-corpus overlaps on \href{https://github.com/LAION-AI/audio-dataset/blob/main/laion-audio-630k/ICASSP.md}{GitHub}. To mitigate any such leakage at evaluation time, we score only on the official held-out test/eval split of each benchmark.}
The emotion portion combines DailyTalk, MELD, EmoVDB, IEMOCAP (Sessions 1--4; Session 5 is held out for evaluation by convention), RAVDESS, and MUStARD++.

\begin{table}[!b]
\centering
\small
\setlength{\tabcolsep}{4pt}
\resizebox{\columnwidth}{!}{%
\begin{tabular}{lrrrr}
\toprule
Modality & Hours & Rows & Prob. & Epochs @ $100$k \\
\midrule
ASR (English)         & $50.8$k & $15.47$M  & $0.65$ & $0.51\text{--}1.17$ \\
Environmental sound   & $4.31$k & $691.8$k & $0.25$ & $4.41\text{--}10.1$ \\
Emotion               & $51$    & $50.28$k & $0.10$ & $24.3\text{--}55.5$ \\
\midrule
Total                 & $55.1$k & $16.22$M  & $1.00$ & --- \\
\bottomrule
\end{tabular}%
}
\caption{Multi-task training mixture. Hours are total source duration (raw, before the $30$-second per-utterance cap). Probabilities are row-level sampling weights for \texttt{interleave\_datasets}. Epochs at $100$k is the per-modality exposure range across encoders (low end: DAC-VAE global batch $16$; high end: EnCodec global batch $32$ with cutoff $4096$), computed as rows-consumed $\times$ prob $/$ rows.}
\label{tab:datasets}
\end{table}

\begin{table}[!t]
\centering
\small
\setlength{\tabcolsep}{4pt}
\begin{tabular}{ll}
\toprule
\rowcolor{gray!15}\multicolumn{2}{l}{\textbf{\textit{Global}}} \\
Optimizer                & AdamW (FusedAdam) \\
Peak LR                  & $2 \times 10^{-4}$ (align) / $2 \times 10^{-5}$ (FT) \\
Weight decay             & $0.01$ \\
Gradient clip            & $1.0$ \\
LR schedule              & warm-up, stable, decay \\
Warmup                   & $1000$ steps \\
Duration                 & $100{,}000$ steps \\
Precision                & bfloat16 \\
\midrule
\rowcolor{gray!15}\multicolumn{2}{l}{\textbf{\textit{LoRA}}} \\
$r$ / $\alpha$ / dropout & $32$ / $64$ / $0.05$ \\
Modules                  & $\{q,k,v,o,\text{gate},\text{up},\text{down}\}_{\text{proj}}$ \\
\bottomrule
\end{tabular}
\caption{Training hyperparameters. Only the peak learning rate differs between the audio-LM alignment phase (align) and the LoRA fine-tuning phase (FT). Per-encoder batch composition (per-device batch, gradient accumulation, effective global batch) is listed in Table \ref{tab:training-config}.}
\label{tab:hparams}
\end{table}

\paragraph{Streaming pipeline.}
The three modality manifests are interleaved at the row level via Hugging Face \texttt{interleave\_datasets} \cite{HF23} with the sampling weights in Table \ref{tab:datasets} and \texttt{stopping\_strategy=all\_exhausted}, so the emotion split cycles multiple times while the ASR split drains.
Audio is capped at $30$ seconds per utterance and resampled online to each encoder's native sampling rate (Table \ref{tab:encoders}).

\paragraph{Sample construction.}
Each sample is wrapped in the ChatML template \cite{OAI23}: a system prompt, a user turn containing $t_{\text{audio}} = \lfloor n_{\text{samples}} / \text{hop} \rfloor$ copies of the audio placeholder token \texttt{<|audio\_pad|>} bracketed by \texttt{<|audio\_start|>} and \texttt{<|audio\_end|>} \cite{Q2A24} (with the encoder-specific hop in Table \ref{tab:encoders}), a modality-specific instruction (\eg ``Transcribe the audio to text.'' for ASR, ``Describe the audio.'' for sound captioning, ``Identify the emotion.'' for emotion recognition), and an assistant turn containing the ground-truth response followed by the end-of-sequence token.
At forward time the projector output replaces the \texttt{<|audio\_pad|>} embeddings in the LLM input, producing an input sequence whose audio-conditioned region has the same length as the projector output.
The training objective is causal cross-entropy computed only on the response and EOS tokens, with all system, user, audio-placeholder, and instruction tokens marked by the ignore-index of $-100$ \cite{HF23}.

\paragraph{Per-modality coverage.}
Applying the row-level sampling weights ($0.65 / 0.25 / 0.10$) to the rows consumed in Table \ref{tab:training-config} gives the per-modality epoch counts.
For DAC-VAE (the smallest global batch), $\approx 12.2$M rows over the mixture translate to $\approx 0.51$ epoch for the ASR portion, $\approx 4.41$ epochs for environmental sound, and $\approx 24.3$ epochs for emotion; the remaining encoders run at roughly twice this consumption.
The ASR coverage is tight enough that some shards may not be drawn at all, while the emotion coverage is high enough that small subsets like RAVDESS ($1{,}440$ rows) are revisited many times within a single training run.
This is a known coverage-versus-mixing trade-off of row-level interleaving with \texttt{stopping\_strategy=all\_exhausted}, and we leave per-modality early stopping or annealed sampling weights to future work.

\paragraph{Training.}
We train in two phases, an audio-LM alignment phase (align) where only the projector is updated and a LoRA fine-tuning phase (FT) where LoRA adapters are inserted into the LLM.
The two phases share all optimization hyperparameters except the peak learning rate (Table \ref{tab:hparams}), and per-encoder batch composition is listed in Table \ref{tab:training-config}.

\section{Datasets and Evaluation Setup}
\label{app:datasets}

\subsection{Dataset Summary}
\label{app:dataset_summary}

\begin{table*}[t]
\centering
\scriptsize
\setlength{\tabcolsep}{4pt}
\newcolumntype{Y}[1]{>{\hsize=#1\hsize\raggedright\arraybackslash}X}
\begin{tabularx}{\textwidth}{@{}l l Y{1.25} Y{0.75}@{}}
\toprule
\textbf{Dataset} & \textbf{Balance} & \textbf{Classes} & \textbf{Notes} \\
\midrule
\rowcolor{gray!15}\multicolumn{4}{l}{\textbf{\textit{Speech emotion}}} \\
IEMOCAP-S5      & unbal 197--613       & angry / happy / neutral / sad & \\
CREMA-D         & unbal 40--51         & angry / disgust / fearful / happy / neutral / sad & \\
\midrule
\rowcolor{gray!15}\multicolumn{4}{l}{\textbf{\textit{Speech content}}} \\
Speech Commands & balanced (200/cls)   & yes / no / up / down / left / right / on / off / stop / go & keyword spotting \\
AudioMNIST      & balanced (200/cls)   & digit\_0 \ldots digit\_9 & spoken digits \\
\midrule
\rowcolor{gray!15}\multicolumn{4}{l}{\textbf{\textit{Vocal sounds}}} \\
VocalSound      & balanced (200/cls)   & cough / laughter / sigh / sneeze / sniff / throat\_clearing & human vocalization without linguistic content \\
\midrule
\rowcolor{gray!15}\multicolumn{4}{l}{\textbf{\textit{Environmental sound}}} \\
ESC-10          & balanced (40/cls)    & chainsaw / clock\_tick / crackling\_fire / crying\_baby / dog / helicopter / rain / rooster / sea\_waves / sneezing & low-resource (40 samples/class) \\
ESC-50 (5-cat)  & balanced (400/cls)   & animals / natural\_soundscapes / human\_sounds / indoor\_sounds / outdoor\_urban & original 50 classes collapsed into 5 super-categories \\
UrbanSound8K    & balanced (100/cls)   & air\_conditioner / car\_horn / children\_playing / dog\_bark / drilling / engine\_idling / gun\_shot / jackhammer / siren / street\_music & \\
TAU Urban 2020  & balanced (500/cls)   & airport / bus / metro / metro\_station / park / public\_square / shopping\_mall / street\_pedestrian / street\_traffic / tram & acoustic scene classification \\
FSD50K\textsuperscript{$\dagger$}        & unbal (multi-label)  & 200 AudioSet-ontology tags & 4-way MCQA with random distractors ($n{=}10{,}231$) \\
AudioSet (eval)\textsuperscript{$\dagger$} & unbal (multi-label) & 632 AudioSet-ontology tags & 4-way MCQA with random distractors ($n{=}3{,}429$ aligned) \\
\midrule
\rowcolor{gray!15}\multicolumn{4}{l}{\textbf{\textit{Music}}} \\
GTZAN           & unbal 99--100        & blues / classical / country / disco / hiphop / jazz / metal / pop / reggae / rock & \\
NSynth          & unbal 180--200       & bass / brass / flute / guitar / keyboard / mallet / organ / reed / string & instrument family, vocal removed \\
IRMAS           & balanced (200/cls)   & acoustic\_guitar / cello / clarinet / electric\_guitar / flute / organ / piano / saxophone / trumpet / violin & predominant instrument in polyphonic mixtures, voice excluded \\
Medley-solos-DB & balanced (200/cls)   & clarinet / distorted\_electric\_guitar / flute / piano / tenor\_saxophone / trumpet / violin & isolated solo instruments, female singer excluded \\
\bottomrule
\end{tabularx}
\caption{
Dataset summary across speech, vocal, environmental sound, and music domains. \textsuperscript{$\dagger$}FSD50K and AudioSet use 4-way MCQA with random distractors per sample because $K$-way over the full class set is impractical.
}
\label{tab:dataset_summary}
\end{table*}

Table \ref{tab:dataset_summary} summarizes the evaluation datasets used in the analyses (Sections~4.1--4.3), spanning speech emotion, speech content, vocal sounds, environmental sound, and music.
For each dataset we list the per-class sample balance, the class set, and a short note describing any non-obvious filtering applied (\eg the 4-class IEMOCAP subset, the $5$-supercategory collapse of ESC-50).

\subsection{Per-Task Metadata}
\label{app:probe_meta}

\begin{table}[htbp]
\centering
\small
\setlength{\tabcolsep}{4pt}
\resizebox{\columnwidth}{!}{%
\begin{tabular}{l rrcccr}
\toprule
\textbf{Task} & \textbf{N} & \textbf{Cls} & \textbf{Bal} & \textbf{Spch} & \textbf{Dur (s)} & \textbf{Chance} \\
\midrule
\rowcolor{gray!15}\multicolumn{7}{l}{\textbf{\textit{Speech emotion}}} \\
CREMA-D       & $264$     & $6$  & no  & yes & $2.5$       & $16.7$\% \\
IEMOCAP       & $2{,}000$ & $4$  & yes & yes & $4.5$       & $25.0$\% \\
\midrule
\rowcolor{gray!15}\multicolumn{7}{l}{\textbf{\textit{Speech content}}} \\
SpeechCmds    & $2{,}000$ & $10$ & yes & yes & $1.0$       & $10.0$\% \\
AudioMNIST    & $2{,}000$ & $10$ & yes & yes & $0.7$       & $10.0$\% \\
\midrule
\rowcolor{gray!15}\multicolumn{7}{l}{\textbf{\textit{Vocal sounds}}} \\
VocalSound    & $1{,}200$ & $6$  & yes & no  & $4.2$       & $16.7$\% \\
\midrule
\rowcolor{gray!15}\multicolumn{7}{l}{\textbf{\textit{Environmental sound}}} \\
ESC-10        & $400$     & $10$ & yes & no  & $5.0$       & $10.0$\% \\
ESC-50 (5cat) & $2{,}000$ & $5$  & yes & no  & $5.0$       & $20.0$\% \\
UrbanSound8K  & $1{,}000$ & $10$ & yes & no  & $\leq 4.0$ & $10.0$\% \\
TAU           & $5{,}000$ & $10$ & yes & no  & $10.0$     & $10.0$\% \\
FSD50K\textsuperscript{$\dagger$}        & $10{,}231$ & $200$ & no  & no  & $\leq 10$   & $25.0$\% \\
AudioSet\textsuperscript{$\dagger$}      & $3{,}429$  & $632$ & no  & no  & $10.0$     & $25.0$\% \\
\midrule
\rowcolor{gray!15}\multicolumn{7}{l}{\textbf{\textit{Music}}} \\
GTZAN         & $999$     & $10$ & no  & no  & $30.1$      & $10.0$\% \\
NSynth        & $1{,}780$ & $9$  & no  & no  & $4.0$       & $11.1$\% \\
IRMAS         & $2{,}000$ & $10$ & yes & no  & $3.0$       & $10.0$\% \\
MedleySolos   & $1{,}400$ & $7$  & yes & no  & $3.0$       & $14.3$\% \\
\bottomrule
\end{tabular}%
}
\caption{Per-task metadata for the linear-probe vs MCQA evaluation in Table~\ref{tab:audio_benchmark_summary}. N: number of test samples. Cls: number of classes. Bal: class-balanced (yes/no). Spch: speech content (yes/no). Dur: median utterance duration in seconds. Chance: uniform-prior chance accuracy. \textsuperscript{$\dagger$}FSD50K and AudioSet use 4-way MCQA with random distractors per sample (chance $25\%$) because $K$-way over the full class set is impractical.}
\label{tab:audio_benchmark_meta}
\end{table}

The linear-probe vs MCQA results table is placed in the main text (Table~\ref{tab:audio_benchmark_summary}, Section~4.1).
Table~\ref{tab:audio_benchmark_meta} lists the per-task metadata used for that analysis (sample count, class count, balance flag, speech flag, median duration, chance accuracy).

\subsection{Per-Benchmark Results}
\label{app:results}

\providecommand{\nodata}{\textcolor{black!35}{--}}

\begin{table*}[t]
\centering
\small
\renewcommand{\arraystretch}{1.05}
\setlength{\tabcolsep}{4pt}
\begin{tabular}{@{}l l c c c c c@{}}
\toprule
\textbf{Benchmark} & \textbf{Metric} & \textbf{Whisper-Small} & \textbf{Whisper-Tiny} & \textbf{EnCodec} & \textbf{WavTokenizer} & \textbf{DAC-VAE} \\
\midrule
Encoder size       & params  & 88.2\,M   & 8.2\,M    & 7.4\,M  & 8.3\,M  & 27.6\,M \\
Representative ckpt & step   & 94{,}000 & 100{,}000 & 90{,}000 & 100{,}000 & 80{,}000 \\
\cmidrule(l){1-7}
\addlinespace[2pt]
\multicolumn{7}{@{}l}{\textbf{ASR}} \\
\addlinespace[1pt]
LibriSpeech test-clean & WER\,$\downarrow$ & \textbf{2.49} & \underline{2.96} & 6.99 & 13.08 & 6.50 \\
                       & CER\,$\downarrow$ & \textbf{0.97} & \underline{1.19} & 3.37 & 7.00 & 2.98 \\
LibriSpeech test-other & WER\,$\downarrow$ & \textbf{6.19} & \underline{8.58} & 22.62 & 38.41 & 23.39 \\
                       & CER\,$\downarrow$ & \textbf{2.96} & \underline{4.33} & 12.88 & 23.86 & 14.00 \\
GigaSpeech test        & WER\,$\downarrow$ & \textbf{12.47} & \underline{14.98} & 32.80 & 49.35 & 31.48 \\
                       & CER\,$\downarrow$ & \textbf{7.13} & \underline{8.56} & 20.27 & 31.53 & 19.12 \\
\cmidrule(l){1-7}
\addlinespace[2pt]
\multicolumn{7}{@{}l}{\textbf{Emotion classification}} \\
\addlinespace[1pt]
MELD       & acc      & \underline{0.483} & \textbf{0.493} & 0.431 & 0.411 & 0.439 \\
           & macro-F1 & \underline{0.287} & \textbf{0.299} & 0.227 & 0.207 & 0.234 \\
IEMOCAP-S5 & acc      & \textbf{0.633} & \underline{0.584} & 0.555 & 0.503 & 0.562 \\
           & macro-F1 & \textbf{0.636} & \underline{0.589} & 0.558 & 0.505 & 0.566 \\
\cmidrule(l){1-7}
\addlinespace[2pt]
\multicolumn{7}{@{}l}{\textbf{Sound captioning}} \\
\addlinespace[1pt]
FSD50K         & BLEU-1    & \textbf{0.715} & \underline{0.655} & 0.352 & 0.501 & 0.344 \\
               & BLEU-4    & \textbf{0.621} & \underline{0.545} & 0.076 & 0.342 & 0.072 \\
               & CIDEr     & \textbf{1.993} & \underline{1.620} & 0.047 & 0.787 & 0.050 \\
Clotho         & BLEU-1    & 0.441 & 0.426 & \textbf{0.457} & 0.338 & \underline{0.446} \\
               & BLEU-4    & \textbf{0.086} & 0.071 & \underline{0.074} & 0.037 & 0.072 \\
               & CIDEr     & \textbf{0.186} & \underline{0.152} & 0.145 & 0.071 & 0.130 \\
AudioCaps      & BLEU-1    & \underline{0.436} & 0.396 & 0.357 & 0.352 & \textbf{0.504} \\
               & BLEU-4    & \underline{0.102} & 0.089 & 0.075 & 0.052 & \textbf{0.123} \\
               & CIDEr     & \textbf{0.351} & \underline{0.292} & 0.254 & 0.141 & 0.282 \\
AudioSet       & BLEU-1    & \textbf{0.515} & \underline{0.471} & 0.459 & 0.386 & 0.408 \\
               & BLEU-4    & \textbf{0.336} & \underline{0.286} & 0.273 & 0.173 & 0.205 \\
               & CIDEr     & \textbf{0.856} & \underline{0.674} & 0.588 & 0.302 & 0.479 \\
\bottomrule
\end{tabular}
\caption{Per-benchmark encoder comparison. ASR reports WER / CER (\%, lower is better). All other metrics are higher-is-better. \textbf{Bold} marks the per-row best, \underline{underline} marks the second-best across encoders (computed only over evaluated cells). \nodata{} = not evaluated. FSD50K and AudioSet are evaluated as captioning tasks (prompt \emph{``Describe what you hear in the audio.''}, BLEU / CIDEr scored against AudioSet-ontology descriptions of the ground-truth labels as multi-reference) rather than vocab-parse F1, because caption-trained Stage-1 models emit free-form descriptions and a vocab parser yields artifact $\approx 0$ that does not reflect actual capability. The same prompt and metric pipeline are used for AudioCaps and Clotho, so the four caption rows are directly comparable. MELD uses the official held-out test split. IEMOCAP Session~5 is a leakage-free cross-corpus emotion benchmark (4-class, exc$\to$hap merge, $n=1241$).}
\label{tab:encoder_comparison}
\end{table*}

\paragraph{Evaluation details.}
FSD50K and AudioSet are evaluated as captioning tasks (prompt ``Describe what you hear in the audio.''), with BLEU and CIDEr scored against AudioSet-ontology descriptions of the ground-truth labels as multi-reference, rather than vocab-parse F1: caption-trained alignment-phase models emit free-form descriptions and a vocab parser yields artifact $\approx 0$ that does not reflect actual capability.
The same prompt and metric pipeline are used for AudioCaps and Clotho, so the four caption rows of Table \ref{tab:encoder_comparison} are directly comparable.
MELD uses the official held-out test split.
IEMOCAP Session~5 is a leakage-free cross-corpus emotion benchmark.
The original IEMOCAP labels span ten emotions plus an unlabeled (xxx) bucket; we follow the conventional $4$-class subset of anger, happy, sad, and neutral, dropping the other labels and merging the excited class into happy to expand the under-represented happy class \cite{Yoon18}.
After this filtering, $1{,}241$ test utterances remain in Session~5.

\begin{table*}[t]
\centering
\small
\setlength{\tabcolsep}{4pt}
\begin{tabular}{ll rrr rrr rrr}
\toprule
 & & \multicolumn{3}{c}{\textbf{Whisper-Tiny}}
   & \multicolumn{3}{c}{\textbf{EnCodec}}
   & \multicolumn{3}{c}{\textbf{DAC-VAE}} \\
 & & \multicolumn{3}{c}{\small (FT ckpt $30$k)}
   & \multicolumn{3}{c}{\small (FT ckpt $70$k)}
   & \multicolumn{3}{c}{\small (FT ckpt $40$k)} \\
\cmidrule(lr){3-5}\cmidrule(lr){6-8}\cmidrule(lr){9-11}
\textbf{Task} & \textbf{Metric} & align & FT & $\Delta$ & align & FT & $\Delta$ & align & FT & $\Delta$ \\
\midrule
\multicolumn{11}{@{}l}{\textbf{ASR} (WER\,$\downarrow$; negative $\Delta$ is improvement)} \\
LibriSpeech test-clean & WER & $2.96$  & $2.62$  & $-0.34$ & $6.99$  & $3.72$  & $-3.27$  & $6.50$  & $5.29$  & $-1.21$ \\
LibriSpeech test-other & WER & $8.58$  & $7.21$  & $-1.37$ & $22.62$ & $13.28$ & $-9.34$  & $23.39$ & $20.05$ & $-3.34$ \\
GigaSpeech test        & WER & $14.98$ & $8.34$  & $-6.64$ & $32.80$ & $14.95$ & $-17.85$ & $31.48$ & $24.48$ & $-7.00$ \\
\midrule
\multicolumn{11}{@{}l}{\textbf{Emotion} (macro-F1\,$\uparrow$)} \\
MELD       & macro-F1 & $0.299$ & $0.265$ & $-0.034$ & $0.227$ & $0.217$ & $-0.010$ & $0.234$ & $0.213$ & $-0.021$ \\
IEMOCAP-S5 & macro-F1 & $0.589$ & $0.596$ & $+0.007$ & $0.558$ & $0.588$ & $+0.030$ & $0.566$ & $0.551$ & $-0.015$ \\
\midrule
\multicolumn{11}{@{}l}{\textbf{Captioning} (CIDEr\,$\uparrow$)} \\
FSD50K    & CIDEr & $1.620$ & $0.813$ & $-0.807$ & $0.047$ & $0.631$ & $+0.584$ & $0.050$ & $0.523$ & $+0.473$ \\
Clotho    & CIDEr & $0.152$ & $0.140$ & $-0.012$ & $0.145$ & $0.111$ & $-0.034$ & $0.130$ & $0.093$ & $-0.037$ \\
AudioCaps & CIDEr & $0.292$ & $0.267$ & $-0.025$ & $0.254$ & $0.222$ & $-0.032$ & $0.282$ & $0.251$ & $-0.031$ \\
AudioSet  & CIDEr & $0.674$ & $0.684$ & $+0.010$ & $0.588$ & $0.668$ & $+0.080$ & $0.479$ & $0.507$ & $+0.028$ \\
\bottomrule
\end{tabular}
\caption{Per-task align vs LoRA-fine-tuning (FT) comparison for the three encoders evaluated under FT. align values are reproduced from Table \ref{tab:encoder_comparison}. FT values are evaluated at the per-encoder best-by-rank-sum checkpoint over the 9-metric pool: Whisper-Tiny ckpt $30{,}000$, EnCodec ckpt $70{,}000$, DAC-VAE ckpt $40{,}000$. Family averages of these per-task numbers appear in Table \ref{tab:lora-summary}.}
\label{tab:lora_per_task}
\end{table*}

\section{LoRA Fine-Tuning: Per-Encoder Details}
\label{app:lora_details}

This section expands the per-encoder case studies behind the summary in Table \ref{tab:lora-summary}; per-task numbers are in Table \ref{tab:lora_per_task}.
On fine-tuned Whisper-Tiny, where align was already strong on non-ASR audio tasks, the LoRA contribution stays close to zero on most non-ASR metrics yet FSD50K captioning regresses sharply, consistent with LoRA over-fitting to the ASR-supervised path and pulling the model away from the already-strong align caption distribution.
On fine-tuned EnCodec, where align had been weak across the board, LoRA produces substantial gains in the headroom-rich slots, while still failing to improve AudioCaps or Clotho, two captioning tasks on which all encoders share a uniformly low ceiling.
Fine-tuned DAC-VAE shows the same pattern, with LoRA again hurting AudioCaps and Clotho captioning and yielding a large FSD50K recovery comparable in scale to EnCodec, while emotion changes remain small.
Across all three encoders, FSD50K is the clearest single-task signature of alignment potential. The LoRA contribution flips from a large drop on the encoder with the strongest align baseline to large gains on the two encoders with the lowest baselines, tracking the room above each encoder's align value.

\paragraph{Probe vs MCQA gap reduction (Whisper-Tiny).}
\label{app:lora_probe_mcqa_gap}

\begin{table*}[t]
\centering
\small
\setlength{\tabcolsep}{6pt}
\begin{tabular}{l rrr rrr r}
\toprule
 & \multicolumn{3}{c}{\textbf{align}}
 & \multicolumn{3}{c}{\textbf{FT}}
 & \\
\cmidrule(lr){2-4}\cmidrule(lr){5-7}
\textbf{Task} & Probe & MCQA & Gap & Probe & MCQA & Gap & $\Delta$Gap \\
\midrule
IEMOCAP      & $56.5$ & $79.5$ & $-23.0$ & $56.0$ & $77.0$ & $-21.0$ & $+2.0$ \\
AudioMNIST   & $99.5$ & $84.0$ & $+15.5$ & $100.0$ & $91.5$ & $+8.5$  & $\mathbf{-7.0}$ \\
VocalSound   & $80.3$ & $43.9$ & $+36.3$ & $77.2$ & $48.5$ & $+28.7$ & $\mathbf{-7.6}$ \\
UrbanSound8K & $74.5$ & $26.0$ & $+48.5$ & $76.0$ & $32.0$ & $+44.0$ & $\mathbf{-4.5}$ \\
GTZAN        & $72.5$ & $8.5$  & $+64.0$ & $71.0$ & $20.5$ & $+50.5$ & $\mathbf{-13.5}$ \\
NSynth       & $67.7$ & $5.6$  & $+62.1$ & $69.2$ & $17.2$ & $+52.0$ & $\mathbf{-10.1}$ \\
\bottomrule
\end{tabular}
\caption{Probe at $L31$ vs MCQA accuracy (\%) for Whisper-Tiny under align and fine-tuning, with the probe-to-MCQA gap (Probe $-$ MCQA) and the change in gap ($\Delta$Gap $=$ Gap$_{\text{FT}}-$Gap$_{\text{align}}$). Negative $\Delta$Gap (bold) marks gap reduction.}
\label{tab:probe_mcqa_gap_wt}
\end{table*}

Table \ref{tab:probe_mcqa_gap_wt} contrasts the probe accuracy at $L31$ with the MCQA accuracy on Whisper-Tiny under align and fine-tuning, on six representative tasks.
Probe accuracy at the final LM layer barely moves between align and fine-tuning (within $\pm 1.5$\,pp on five of six tasks, $-3.1$\,pp on VocalSound), indicating that LoRA does not noticeably reshape the LM's internal acoustic representation.
MCQA accuracy, by contrast, rises on every non-emotion task, with the largest absolute gains where the LM previously failed to convert latent acoustic evidence into the correct letter.
As a result, the probe-to-MCQA gap shrinks across all five non-emotion tasks, most strongly on GTZAN and NSynth, with smaller reductions on VocalSound, AudioMNIST, and UrbanSound8K. IEMOCAP is the exception, where MCQA already exceeded the probe at $L31$ under align (Gap $=-23.0$), leaving little remaining gap for fine-tuning to close.

\section{Readout-Level Diagnostics}
\label{app:readout}

\subsection{LogitLens across All Tasks}
\label{app:logit_lens_grid}

\begin{figure*}[htbp]
  \centering
  \includegraphics[width=\textwidth, trim=0 153.9 0 0, clip]{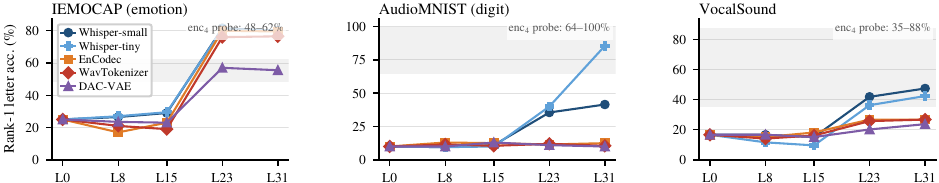}

  \vspace{6pt}
  \includegraphics[width=\textwidth, trim=0 0 0 140.0, clip]{figures/causal_logit_lens_grid.pdf}
  \caption{LogitLens rank-1 letter accuracy across all six tasks, decoded at five LM depths ($L0, L8, L15, L23, L31$ of the 32-layer Qwen3.5-4B) via final RMSNorm $+\,W_U$ and restricted to the candidate-letter tokens. Each line is one encoder family; the grey band marks the range of $E4$ encoder-probe accuracies across the five families for that task. Speech tasks reach a high $L31$ accuracy that tracks audio-on MCQA, while music tasks and UrbanSound8K stay near chance at every depth despite high probe accuracy, isolating the readout as the bottleneck.}
  \label{fig:logit_lens_grid}
\end{figure*}

\begin{figure*}[htbp]
  \centering
  \includegraphics[width=\textwidth, trim=0 162.6 0 0, clip]{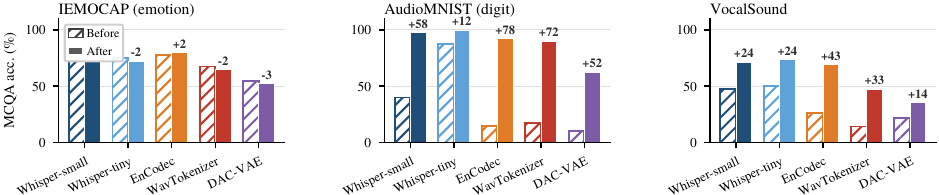}

  \vspace{6pt}
  \includegraphics[width=\textwidth, trim=0 0 0 159.7, clip]{figures/causal_recovery_grid.pdf}
  \caption{Per-task $W_U$ letter-row fine-tune recovery, the task $\times$ encoder breakdown of Figure~\ref{fig:causal_verify}. Hatched bar: MCQA accuracy before fine-tune (equivalent to the rank-1 letter accuracy from final-layer LogitLens). Filled bar: MCQA accuracy after fine-tuning only the $n_c$ rows of $W_U$ that index the choice letters, with everything else frozen; the $\pm$ label above each After bar is the absolute change in pp. IEMOCAP shows near-zero recovery because the readout is already aligned; the other five tasks show large recoveries up to $+78$\,pp, with the gain pattern across encoders mirroring the LogitLens curves in Figure~\ref{fig:logit_lens_grid}.}
  \label{fig:recovery_grid}
\end{figure*}

Figure~\ref{fig:logit_lens_grid} applies a LogitLens protocol across the six analysis tasks: we cache the assistant-newline hidden state at five depths ($L0, L8, L15, L23, L31$ of the 32-layer Qwen3.5-4B), apply the LM's own final RMSNorm followed by $W_U$, and read logits restricted to the candidate-letter tokens; the curves are per-encoder rank-1 accuracy, and the shaded band marks the range of encoder-stage probe accuracies at $E4$ across the five families.

The grid sharpens the speech--music split already visible in the rank-delta heatmap (Figure~\ref{fig:rank_delta_heatmap}).
On speech-centric tasks (IEMOCAP, AudioMNIST, and to a lesser extent VocalSound) the LogitLens climbs through the LM and saturates at $L31$ to a level comparable to the corresponding audio-on MCQA accuracy, confirming that the readout is faithful when the LM has actually learned to route acoustic evidence to the letter space.
On music-centric tasks (NSynth, GTZAN) and the noisy environmental task UrbanSound8K, the trajectory stays flat near chance at every depth, even though the encoder-probe band sits well above (e.g.\ $38$--$97\%$ for NSynth, $47$--$86\%$ for GTZAN).
The gap between the probe band and the saturated LogitLens curve at $L31$ is therefore a per-task estimate of how much of the available acoustic structure $W_U$ fails to convert into a correct letter, and it is largest exactly on the tasks where the audio effect (Figure~\ref{fig:rank_delta_heatmap}) is near zero or negative.

\subsection{\texorpdfstring{$W_U$}{W\_U} Letter-Row Fine-Tune by Task}
\label{app:wu_finetune_grid}

The main-text figure (Figure~\ref{fig:causal_verify}) reports $W_U$ letter-row fine-tune recovery on 6 tasks, which gives a concise from $+30$ to $+42$\,pp range but masks two things: (i) on tasks where the original readout is already aligned, the per-encoder recovery is near zero, and (ii) on the tasks where the readout is misaligned, the per-encoder recovery is substantially larger than the cross-task mean.
Figure~\ref{fig:recovery_grid} breaks the same experiment down by task $\times$ encoder family.
The protocol is identical to Section~4.3: encoder, projector, and all LM weights are frozen, the $n_c$ rows of $W_U$ corresponding to the choice letters are trained with cross-entropy on saved $L31$ assistant-newline hidden states for $200$ full-batch Adam steps at lr $10^{-3}$, with an 80/20 stratified split (test accuracy reported).

The per-task panels match the LogitLens grid (Figure~\ref{fig:logit_lens_grid}) one-for-one.
IEMOCAP-emotion sits in the regime where $W_U$ is already approximately aligned with the class-mean directions: pre-fine-tune accuracy is already in the $55$--$78\%$ band across encoders, and fine-tuning the letter rows changes accuracy by at most a couple of points (in some cases negatively, consistent with mild overfitting on $\sim$$160$ training samples).
The remaining five tasks all show the readout-bottleneck regime where the letter rows are misaligned and a tiny surgical update is enough to recover most of the gap.
The largest task-specific deltas reach $+78$\,pp (AudioMNIST/EnCodec), $+58$\,pp (NSynth/Whisper-Small), and $+58$\,pp (GTZAN/EnCodec), each measured against a base accuracy below $20\%$ and recovered using only $n_c \times 2560$ scalars (between $\sim$$10\text{k}$ and $\sim$$26\text{k}$ depending on the task).
Because the entire LM, projector, and encoder are frozen, these gains cannot reflect new representation learning; they directly quantify how much MCQA accuracy is held back by misalignment between class-mean directions and the letter rows of $W_U$.

\subsection{Logit Processor Baseline: Vocabulary Restriction Alone Is Insufficient}
\label{app:logit_processor}

The $W_U$ letter-row fine-tune in Appendix~\ref{app:wu_finetune_grid} recovers from $+30$ to $+78$\,pp of MCQA accuracy, but this leaves open whether the gain comes from (i) actually re-aligning $W_U$ to the class-mean directions or (ii) merely keeping the argmax inside the candidate-letter set $\{A,\dots,K\}$, which is essentially equivalent to vocabulary masking at decode time.
We separate the two effects by treating the vocabulary restriction as its own intervention.
For every (task, encoder) combination we run greedy decoding under the HuggingFace \texttt{LogitsProcessor} API in three conditions: \textbf{(B0)} vanilla LM, argmax over the full $248{,}320$-token vocabulary; \textbf{(B1)} same LM, argmax over the candidate-letter token set only (non-letter logits set to $-\infty$); and \textbf{(B2)} the $W_U$ letter-row fine-tune of Appendix~\ref{app:wu_finetune_grid}.

\begin{figure}[t]
  \centering
  \includegraphics[width=\columnwidth]{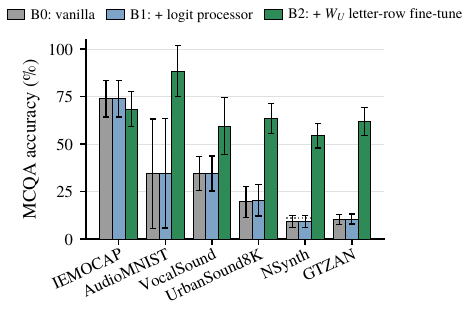}
  \caption{MCQA accuracy across three readout interventions, averaged over five encoders. \textbf{B0}: vanilla LM, full-vocab argmax. \textbf{B1}: + Logit processor restricting argmax to $\{A,\dots,K\}$. \textbf{B2}: + $W_U$ letter-row fine-tune (Appendix~\ref{app:wu_finetune_grid}). Error bars: $\pm$1\,SD across encoders. Dotted segment over each group: per-task chance. Across $30$ cells, B1 changes B0 accuracy by only $+0.23$\,pp on average, while B2 recovers $+25$--$+54$\,pp on the five readout-bottleneck tasks; IEMOCAP shows a small negative B2 effect (mild overfitting on the $\sim$$162$ training samples; its readout is already approximately aligned).}
  \label{fig:logit_processor}
\end{figure}

Across all $30$ combinations (5 encoders $\times$ 6 tasks), B1 changes B0 accuracy by an average of only $+0.23$\,pp (max $+1.0$\,pp on EnCodec/GTZAN); the model's emission rate to candidate-letter tokens averages $98.5\%$ and never falls below $94\%$, so the vocabulary mask is effectively redundant.
Only the $W_U$ letter-row update closes the readout gap (Figure~\ref{fig:logit_processor}).
This rules out the possibility that the surgical recovery merely reflects ``forcing a letter answer'': the failure is letter-vs-letter ranking \emph{inside} the candidate set, not non-letter token emission.

\begin{table}[htbp]
\centering
\small
\renewcommand{\arraystretch}{1.05}
\setlength{\tabcolsep}{4pt}
\resizebox{\columnwidth}{!}{%
\begin{tabular}{@{}l l c c@{}}
\toprule
\textbf{Benchmark} & \textbf{Metric} & \textbf{Whisper-Small} & \textbf{DAC-VAE} \\
\midrule
Encoder size        & params & 88.2\,M   & 27.7\,M \\
Representative ckpt & step   & 90{,}000  & 90{,}000 \\
\cmidrule(l){1-4}
\addlinespace[2pt]
\multicolumn{4}{@{}l}{\textbf{ASR}} \\
\addlinespace[1pt]
LibriSpeech test-clean & WER\,$\downarrow$ & \textbf{4.83}  & 8.63 \\
                       & CER\,$\downarrow$ & \textbf{2.67}  & 5.47 \\
LibriSpeech test-other & WER\,$\downarrow$ & \textbf{8.21}  & 19.77 \\
                       & CER\,$\downarrow$ & \textbf{4.38}  & 11.74 \\
GigaSpeech test        & WER\,$\downarrow$ & \textbf{11.51} & 20.02 \\
                       & CER\,$\downarrow$ & \textbf{5.73}  & 11.10 \\
\cmidrule(l){1-4}
\addlinespace[2pt]
\multicolumn{4}{@{}l}{\textbf{Emotion classification}} \\
\addlinespace[1pt]
MELD       & acc      & 0.111 & \textbf{0.196} \\
           & macro-F1 & 0.132 & \textbf{0.159} \\
IEMOCAP-S5 & acc      & \textbf{0.383} & 0.344 \\
           & macro-F1 & \textbf{0.479} & 0.442 \\
\cmidrule(l){1-4}
\addlinespace[2pt]
\multicolumn{4}{@{}l}{\textbf{Sound captioning}} \\
\addlinespace[1pt]
FSD50K    & BLEU-1 & \textbf{0.223} & 0.203 \\
          & BLEU-4 & \textbf{0.116} & 0.091 \\
          & CIDEr  & 0.118 & 0.118 \\
Clotho    & BLEU-1 & \textbf{0.195} & 0.171 \\
          & BLEU-4 & \textbf{0.028} & 0.024 \\
          & CIDEr  & \textbf{0.113} & 0.059 \\
AudioCaps & BLEU-1 & 0.180 & \textbf{0.331} \\
          & BLEU-4 & 0.033 & \textbf{0.068} \\
          & CIDEr  & \textbf{0.262} & 0.242 \\
AudioSet  & BLEU-1 & \textbf{0.235} & 0.201 \\
          & BLEU-4 & \textbf{0.114} & 0.081 \\
          & CIDEr  & \textbf{0.054} & 0.037 \\
\bottomrule
\end{tabular}%
}
\caption{Downstream benchmark performance of the two encoders paired with Ministral3-3B (Instruct), the second base LM. Best value per row in bold. Values are reported on the scale used in Table~\ref{tab:encoder_comparison}: WER and CER as percentages, all other metrics in $[0,1]$.}
\label{tab:ministral-benchmark}
\end{table}

\begin{table}[htbp]
\centering
\small
\setlength{\tabcolsep}{4pt}
\resizebox{\columnwidth}{!}{%
\begin{tabular}{l ccc ccc}
\toprule
\multirow{2}{*}{\textbf{Task}}
  & \multicolumn{3}{c}{\textbf{Whisper-Small}}
  & \multicolumn{3}{c}{\textbf{DAC-VAE}} \\
\cmidrule(lr){2-4}\cmidrule(lr){5-7}
  & MCQA & $W_U$ FT & Probe & MCQA & $W_U$ FT & Probe \\
\midrule
AudioMNIST    & $76.8$ & $99.9$ & $100.0$ & $10.7$ & $98.5$ & $98.8$ \\
NSynth        & $8.3$  & $62.4$ & $63.8$  & $11.6$ & $63.1$ & $63.1$ \\
GTZAN         & $10.0$ & $77.9$ & $78.6$  & $10.0$ & $75.6$ & $76.9$ \\
UrbanSound8K  & $25.2$ & $82.8$ & $83.0$  & $12.0$ & $73.3$ & $76.5$ \\
IEMOCAP       & $58.8$ & $65.8$ & $65.6$  & $50.9$ & $60.3$ & $60.3$ \\
VocalSound    & $33.7$ & $88.8$ & $89.7$  & $15.5$ & $73.3$ & $74.3$ \\
\midrule
\textit{Average} & $35.5$ & $79.6$ & $80.1$ & $18.5$ & $74.0$ & $75.0$ \\
\bottomrule
\end{tabular}%
}
\caption{Readout bottleneck under a second base LM. MCQA accuracy, $W_U$ letter-row fine-tuning, and final-layer linear probe accuracy for two encoders paired with Ministral3-3B (Instruct).}
\label{tab:ministral-readout}
\end{table}

\subsection{Generality of the Readout Bottleneck across Base LMs}
\label{app:second_lm}

To test whether the readout bottleneck is specific to Qwen3.5-4B, we repeat the analysis with Ministral3-3B (Instruct) as the base LM, keeping the encoder, projector, and training protocol otherwise unchanged.
Table~\ref{tab:ministral-benchmark} reports downstream benchmark performance for Whisper-Small and DAC-VAE, which differ substantially in end-task strength, most visibly on ASR.
Table~\ref{tab:ministral-readout} reports MCQA, $W_U$ letter-row fine-tuning, and linear-probe accuracy for the same two encoders.
The letter-row update recovers accuracy to within about a point of the probe on every task while direct MCQA stays far below both, matching the pattern reported for Qwen3.5-4B.
Both encoders show this gap despite their difference in end-task strength, so the bottleneck is not specific to a single LM family and does not follow from a particular level of downstream capability.

\subsection{Robustness of the Readout Intervention}
\label{app:readout_robustness}

\paragraph{Protocol.}
The ablations here use each task's own standard train/test split (Table~\ref{tab:wu-intervention-protocol}): the letter rows are fit on the training split, and every accuracy is reported on the disjoint test split, with original MCQA and the post-intervention score measured on the same examples. Training uses AdamW (lr $5\times10^{-4}$, weight decay $10^{-3}$), mini-batch $128$, $150$ epochs, with gradients masked to the $n_c$ candidate-letter rows and all other parameters frozen.

\begin{table}[htbp]
\centering
\small
\setlength{\tabcolsep}{4pt}
\begin{tabular}{@{}l c r r@{}}
\toprule
\textbf{Task} & \textbf{$n_c$} & \textbf{Train} & \textbf{Split protocol} \\
\midrule
AudioMNIST   & 10 & 2{,}000 & native train/test \\
NSynth       & 10 & 2{,}000 & native train/test \\
GTZAN        & 10 & 700     & seed-fixed disjoint \\
UrbanSound8K & 10 & 2{,}000 & official folds \\
IEMOCAP      & 4  & 5{,}370 & session held-out \\
VocalSound   & 6  & 1{,}200 & native train/test \\
\bottomrule
\end{tabular}
\caption{Data protocol for the readout-robustness ablations (Tables~\ref{tab:mcqa-format-variants} and~\ref{tab:readout-conditions}). Each task uses its full standard training split, distinct from the $200$-per-class evaluation subsample of Table~\ref{tab:dataset_summary}, and the reported accuracy is measured on that split's evaluation partition, which is disjoint from the states used to fit the letter rows. Original MCQA and post-intervention accuracy are paired measurements on the same evaluation examples, computed from the same saved $L31$ hidden states. Only the $n_c$ rows of $W_U$ indexing the choice letters receive gradient, giving $n_c \times d_{\text{model}}$ updated scalars per task, that is $10$--$26$k for Qwen3.5-4B ($d_{\text{model}}{=}2560$) and up to $31$k for Ministral3-3B ($d_{\text{model}}{=}3072$); every other parameter, including the embedding matrix and the remaining vocabulary rows, stays frozen at its pretrained value.}
\label{tab:wu-intervention-protocol}
\end{table}

\begin{table*}[htbp]
\centering
\small
\setlength{\tabcolsep}{5pt}
\begin{tabular}{l ccccccc}
\toprule
\textbf{Condition} & \textbf{AudioMNIST} & \textbf{NSynth} & \textbf{GTZAN} & \textbf{UrbanSound8K} & \textbf{IEMOCAP} & \textbf{VocalSound} & \textbf{Average} \\
\midrule
Baseline      & $74.5$ / $99.6$  & $14.7$ / $56.5$ & $13.7$ / $76.6$ & $35.2$ / $81.0$ & $57.9$ / $63.1$ & $54.0$ / $87.3$ & $41.7$ / $77.4$ \\
Letter-remap  & $76.8$ / $99.2$  & $11.8$ / $55.8$ & $9.0$ / $74.6$  & $16.7$ / $81.3$ & $53.4$ / $61.8$ & $35.0$ / $86.8$ & $33.8$ / $76.6$ \\
Order-shuffle & $71.1$ / $99.8$  & $13.0$ / $55.6$ & $13.0$ / $73.6$ & $34.8$ / $80.3$ & $58.9$ / $62.3$ & $49.8$ / $86.5$ & $40.1$ / $76.4$ \\
Instr-C1      & $63.8$ / $99.7$  & $16.4$ / $57.6$ & $14.4$ / $73.6$ & $34.2$ / $82.5$ & $56.9$ / $62.2$ & $54.0$ / $87.5$ & $40.0$ / $77.2$ \\
Instr-C2      & $64.3$ / $99.7$  & $17.6$ / $55.3$ & $14.7$ / $74.6$ & $36.3$ / $81.5$ & $56.1$ / $62.3$ & $52.2$ / $87.3$ & $40.2$ / $76.8$ \\
Label-word    & $99.0$ / $100.0$ & $12.1$ / $58.9$ & $17.1$ / $75.6$ & $38.0$ / $82.8$ & $51.2$ / $58.3$ & $64.5$ / $88.8$ & $47.0$ / $77.4$ \\
\bottomrule
\end{tabular}
\caption{Robustness of the $W_U$ letter-row intervention across MCQA answer formats (Whisper-Small). Each cell reports original MCQA accuracy / accuracy after the letter-row fine-tune (\%). \textit{Letter-remap} changes the choice-letter mapping, \textit{Order-shuffle} permutes the option order, \textit{Instr-C1} uses the instruction ``Listen to the audio and answer.'' and \textit{Instr-C2} uses ``You are given an audio clip. Choose the option that best describes the sound you hear. Respond with only the option identifier.'', and \textit{Label-word} replaces each choice letter with its candidate label word (\eg ``A. piano'' becomes ``piano''), scored by the first-token logit among the candidate label words and verified collision-free across the six task vocabularies.}
\label{tab:mcqa-format-variants}
\end{table*}

\begin{table*}[htbp]
\centering
\small
\setlength{\tabcolsep}{5pt}
\resizebox{\linewidth}{!}{%
\begin{tabular}{ll cccccc}
\toprule
\textbf{Method} & \textbf{Intervention} & \textbf{IEMOCAP} & \textbf{AudioMNIST} & \textbf{VocalSound} & \textbf{UrbanSound8K} & \textbf{NSynth} & \textbf{GTZAN} \\
\midrule
Original MCQA           & --                                   & $57.9$ & $74.5$ & $54.0$ & $35.2$ & $14.7$ & $13.7$ \\
Bias-only calibration   & Bias update only                     & $60.7$ & $79.0$ & $54.0$ & $40.2$ & $16.7$ & $14.0$ \\
Letter-row fine-tune    & Direction learning (letter rows)     & $64.7$ & $99.7$ & $87.3$ & $82.2$ & $59.7$ & $74.9$ \\
Full-row update         & Direction learning (full vocabulary) & $64.0$ & $99.7$ & $87.3$ & $82.2$ & $59.0$ & $74.9$ \\
\midrule
Final-layer linear probe & Supervised classifier               & $64.3$ & $99.8$ & $87.7$ & $82.7$ & $59.6$ & $79.6$ \\
\bottomrule
\end{tabular}%
}
\caption{Readout condition ablation (Whisper-Small, MCQA accuracy in \%, same train/test split throughout). The probe row is an upper reference rather than an intervention. It fits an unconstrained classifier and does not operate through $W_U$.}
\label{tab:readout-conditions}
\end{table*}

Table~\ref{tab:mcqa-format-variants} varies the prompt template, option order, choice-letter mapping, and label verbalization while holding the rest of the protocol fixed.
Post-intervention accuracy stays within roughly one point of the baseline condition in every variant, including the condition that replaces choice letters with semantic label words, so the recovery is not tied to a single prompt or answer-token format.
AudioMNIST is an expected exception under that condition, where spelling the digits out nearly saturates accuracy before any intervention, consistent with Whisper's ASR pretraining.

Table~\ref{tab:readout-conditions} separates what the intervention corrects.
Calibrating the answer-token biases alone recovers little, while learning a readout direction for the letter rows recovers most of the gap and lands close to an unconstrained final-layer linear probe. Extending the update to the full vocabulary adds nothing further.
The gains therefore reflect a corrected mapping from the final-layer representation to the answer tokens rather than bias correction, and the small residual gap to the probe indicates that most recoverable structure is already present at the final hidden state.

\end{document}